# Long Term Sunspot Cycle Phase Coherence with Periodic Phase Disruptions

*Gerald E. Pease, Gregory S. Glenn*

## Abstract

In 1965 Paul D. Jose published his discovery that both the motion of the Sun about the center of mass of the solar system and periods comprised of eight Hale magnetic sunspot cycles with a mean period of ~22.37 years have a matching periodicity of ~179 years. We have investigated the implied link between solar barycentric torque cycles and sunspot cycles and have found that the unsigned solar torque values from 1610 to 2057 are consistently phase and magnitude coherent in ~179 year Jose Cycles. We are able to show that there is also a surprisingly high degree of sunspot cycle phase coherence for times of minima in addition to magnitude correlation of peaks between the nine Schwabe sunspot cycles of 1878.8 to 1976.1 (SC12 through SC20) and those of 1699 to 1798.3 (SC[-5] through SC4). We further show that the remaining seven Schwabe cycles in each ~179 year cycle are non-coherent. In addition we have analyzed the empirical solar motion triggers of both sunspot cycle phase coherence and phase disruption, from which we conclude that sunspot cycles SC28 through SC35 (2057 to 2143) will be phase coherent at times of minima and amplitude correlated at maxima with SC12 through SC19 (1878.8-1964.8). The resulting predicted start times ± 0.9 year, 1 sigma, of future sunspot cycles SC28 to SC36 are tabulated.

*Keywords*: Jose Cycle barycentric torque subcycle SILSO Jupiter Saturn Uranus synodic resonance syzygies JPL DE405 solar ephemeris planetary dynamo Dalton Galileo Maunder Minimum minima phase jitter tachocline helicity Waldmeier

## 1.0 Preliminary Analysis

Preliminary to performing our analysis, we generated relatively high resolution (5-day step size) plots of solar angular momentum, L (Figure 1) and torque, T = dL/dt (Figure 2). These were computed from the CalTech JPL Horizons DE405 barycentric solar ephemeris website, http://ssd.jpl.nasa.gov/horizons.cgi#top in order to examine the magnitudes of planetary perturbation forces on the Sun.

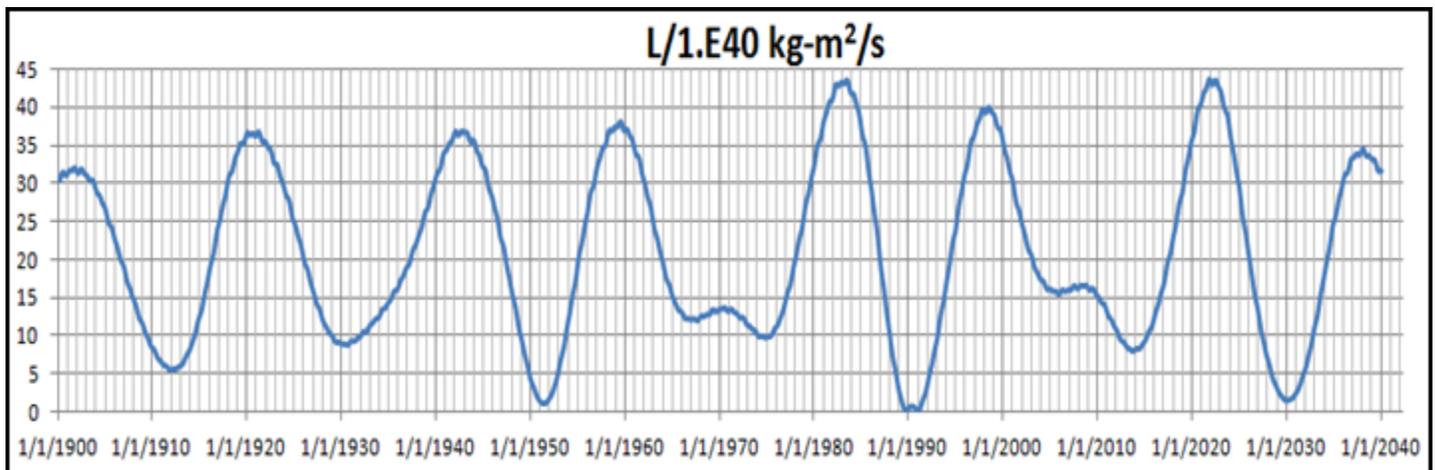

**Figure 1.** Barycentric solar angular momentum, $L = M_{sun}((Y\dot{Z}-Z\dot{Y})^2+(Z\dot{X}-X\dot{Z})^2+(X\dot{Y}-Y\dot{X})^2)^{1/2}$

The primary torque cycles are produced by Jupiter and Saturn. In Figure 2, note also the 1967-1974 and 2004-2010 phase distortions from Uranus' non-resonant low frequency harmonics. Neptune is Jose Cycle phase coherent with the undistorted torque cycles of Jupiter. Venus, Earth, and Mercury produce the vertical spikes.

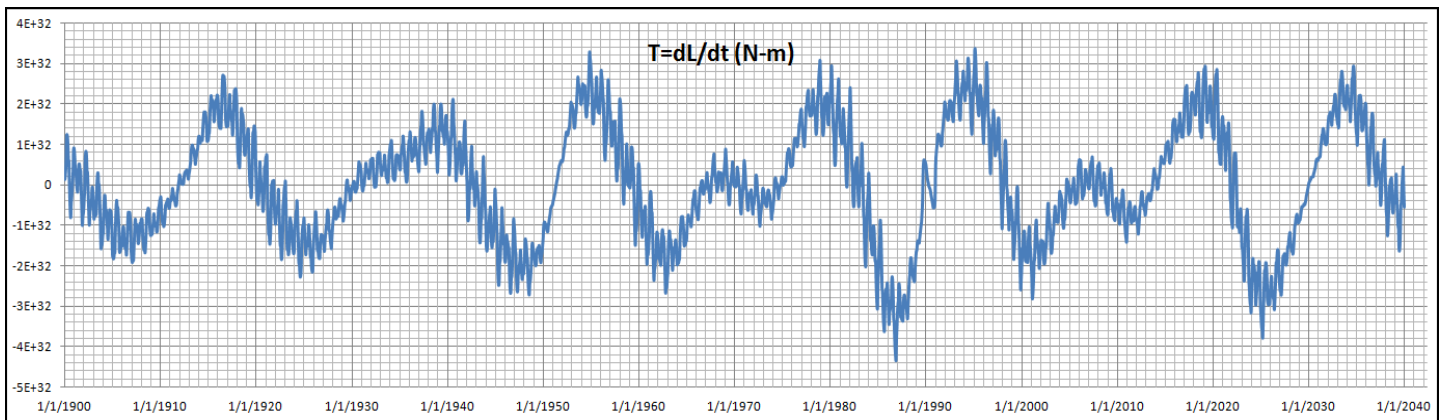

**Figure 2.** Barycentric solar torque T = dL/dt (N-m), 1900-2040

Figure 3 shows that all of the barycentric solar torque cycles in the 179 year Jose Cycle 1878-2057 (red) are surprisingly phase and magnitude coherent with the torque cycles in the previous Jose Cycle 1699-1878 (blue). Annual absolute torque values, normalized for later comparison with sunspot cycles in Figure 5 are plotted.

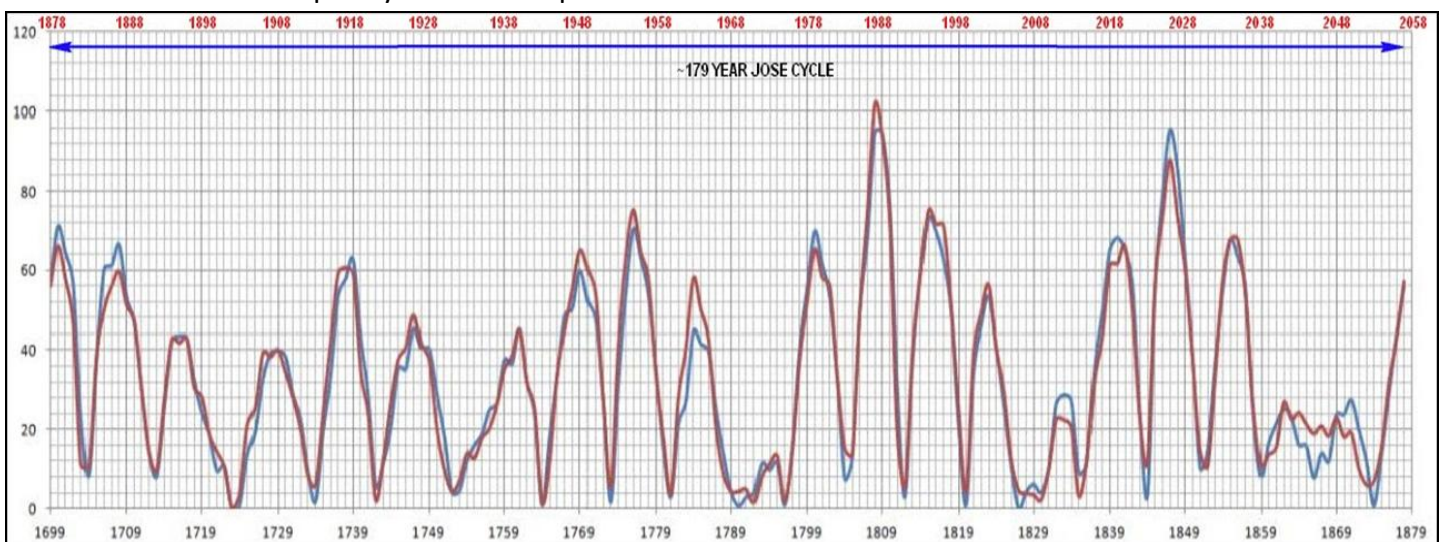

**Figure 3.** Normalized annual absolute values of barycentric torque, 1699-1878 (blue) and 1878-2057 (red).

As will be seen later in Figures 7 to 34, the Jose Cycle torque cycle coherence causes a fairly precise repetition of the complex path of the Sun at ~179 year intervals. In addition to the Figure 2 1967-1974 and 2004-2010 distortions, a 2038-2053 declining torque plateau is clearly seen on Figure 3.

Figure 4 compares the 1699-1878 mean annual SILSO V1.0 Wolf sunspot numbers (blue) with 1878-2057 (projected) values (red). The nine sunspot cycles of 1878-1976 (SC12 through SC20) are seen to display a high degree of phase coherence and amplitude correlation with SC[-5] through SC20 (1699-1798), along with an apparent amplitude reinforcement of sunspot cycles SC17 through SC19 1933-1964). Recent WDC-SILSO V2.0 monthly smoothed SSN adjustments include Waldmeier weighting bias corrections, and respectively decrease the Figure 4 SILSO v1.0 SC18 and SC19 amplitudes relative to SC2 and SC3 by 29.5% and 5.4%. The weighting bias issue is explained in https://arxiv.org/ftp/arxiv/papers/1507/1507.01119.pdf).

SC[-4] appears to be ~+1.5 years offset in phase and SC[-3] is a +100% amplitude mismatch with SC14. In Figure 4, note also the 180° phase difference maintained from 1989 to 2012 and the possibly associated destructive amplitude effects.

Figure 4 also shows our projection of an SC24 sunspot minimum in late 2019 and a rough projection of SC25, SC26, and SC27 for a scenario in which the end of SC27 becomes phase coherent with the end of SC11.

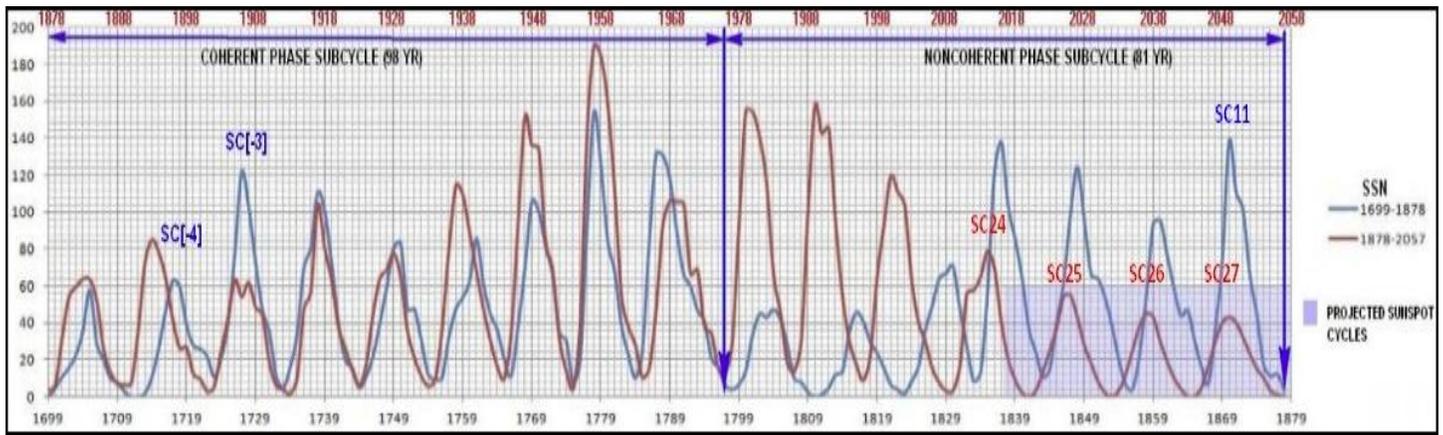

**Figure 4.** SILSO V1.0 Sunspot cycle phase and amplitude comparison of 1878-2057 Jose Cycle with 1699-1878 Jose Cycle.

Table 1 uses more precise sunspot minima times from WDC-SILSO V2, www.solen.info/solar/cycles1_to_present.

| Sunspot Cycles | SC1,SC17 | SC2,SC18 | SC3,SC19 | SC4,SC20 | SC5,SC21 | Mean | StdDev |
|---|---|---|---|---|---|---|---|
| Start (SC1-SC5) | 1755.2 | 1766.4 | 1775.5 | 1784.7 | 1798.3 | | |
| End (SC1-SC4) | 1766.4 | 1775.5 | 1784.7 | 1798.3 | | | |
| Start (SC17-SC21) | 1933.7 | 1944.3 | 1954.3 | 1964.8 | 1976.2 | | |
| End (SC17-SC20) | 1944.3 | 1954.3 | 1964.8 | 1976.1 | | | |
| Derived Cycle(yrs) | 178.5 | 177.9 | 178.8 | 180.1 | 177.9 | 178.64 | 0.90 |
| Max SC1-SC4 | 1761.4 | 1769.8 | 1778.3 | 1788 | | | |
| Max SC17-SC20 | 1937.4 | 1947 | 1958 | 1969.2 | | | |
| DerivedMaxCycle | 176.0 | 177.2 | 179.7 | 181.2 | | 178.53 | 2.04 |

**Table 1.** Derived sunspot cycle phase coherence and phase jitter for sunspot minima times known to nearest 0.1 year.

The period 1933.7 to 1976.1 (SC17 through SC20) is sunspot cycle phase coherent with the period 1755.2 to 1798.3 (SC1 through SC4) and, additionally, the start and end times of those eight sunspot cycles are currently known to a relatively high degree of precision (0.1 year).

By differencing the higher precision start and stop times of SC1 through SC4 from those of SC17 through SC20, we have derived in Table 1 a mean observed sunspot phase coherence cycle value of 178.6 years ± 0.9 year corresponding to observed phase jitter. The derived sunspot max cycle values have fewer samples and twice as much scatter as the derived start time phase coherent cycles. 178.6 years and ±0.9 year are therefore considered to be the best estimates of the mean derived coherence cycle value and associated phase jitter. They are consistent with the observed phase differences at minima between 1755.2-1798.3 and 1933.7-1976.2 in Figure 4 and Table 1.

Figure 5 again shows the start years of sunspot cycle phase coherence in 1878 and 1699. Those years are preceded by anomalously short and flat torque cycles in the 15 year intervals 1859 to 1874 and 1680-1695. The corresponding future years are 2038 to 2053.

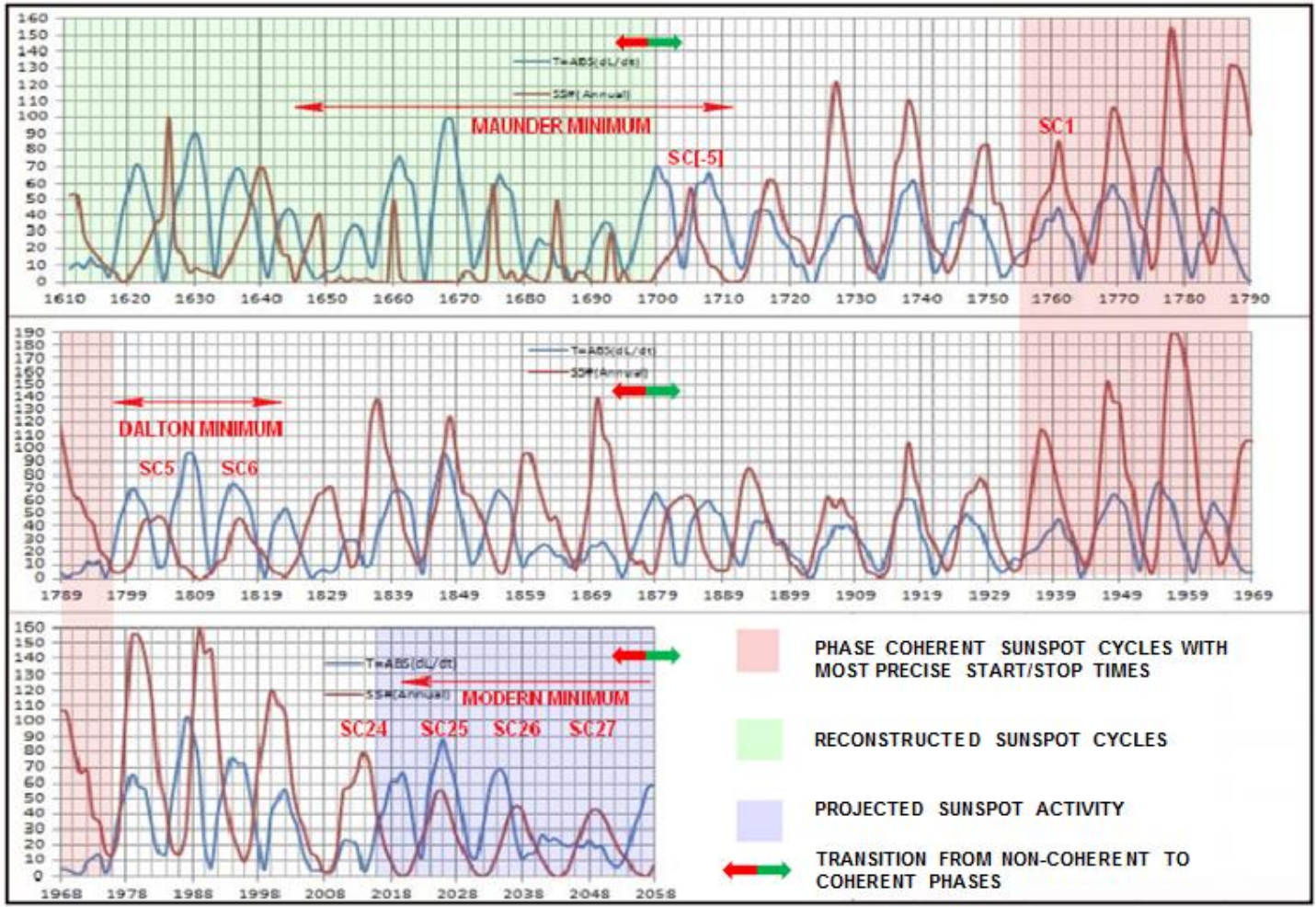

**Figure 5.** Normalized ABS Torque (blue) and SILSO v1.0 Sunspot (red) cycles, 1610 to 2016, with projected sunspot cycles to 2058.

Figure 5 starts in 1610 with a torque plateau and sunspot numbers derived from the telescopic observations of Galileo followed, beginning in 1626, by a combination of attempted sunspot number and sunspot cycle maxima reconstructions by Waldmeier and Schove.

Although Figure 5 projects SC25 through SC27 to have very low magnitudes reminiscent of the Dalton Minimum, the projected SILSO v1.0 Wolf sunspot cycles could turn out to be more (or even less) active than shown, and the slide to coherent phasing could arguably be either slower or faster. Because we are currently in a non-coherent sunspot phase period, the uncertainties of the projected minima times and maxima amplitudes are high.

Figure 5 also shows a possibility of phase coherence of the 1798 and 1823 minima at the start and end of Dalton Minimum cycles SC5 and SC6 with the data-poor reconstructed minima of 1619 and 1645, and shows the phase coherent period previously noted in Figure 4 (1878 to 1976 with 1699 to 1797 ).

The period 1968 to 1976 is sunspot phase coherent with 1789 to 1797 and also with 1610 to 1618 if the uncertainties of sunspot cycle reconstructions for that time period are taken into account. Between 1976.2 (the start of SC21) and 1989 (the peak of SC22) both phase coherence and amplitude correlation with the Dalton Minimum cycles SC5 and SC6 (1798.2 to 1823.4) are broken.

From 1989 to the 2012 first peak of SC24 the sunspot numbers are seen in both Figure 4 and Figure 5 to be 180 degrees out of phase with the sunspot numbers of 1811 through 1834 (SC6 and SC7). $180°$ phase sunspot cycle destructive interference from 1989 to 2012 may be the cause of the reduced magnitudes of SC23 and SC24, and perhaps may also reduce the amplitude of SC25 to slightly less than that of SC24 as projected. Figure 4 shows that a large phase difference will be sufficiently maintained through 2018, which may cause SC24 to reach sunspot minimum by late 2019. The possibility of an abrupt end of SC24 is consistent with the observed rapid downward trend of SC24 in late 2016.

The 16 sunspot cycles in the 179 years from 1699.8 to 1878.8 (SC[-5] through SC11) have a mean length of 11.19 years. The 9 coherent sunspot cycles from 1878.8 to 1976.1 (SC12 through SC20) span 97.3 years and the 9 from 1699.8 to 1798.3 (SC[-5] through SC4) span 98.5 years. The mean phase-coherent sunspot cycle length corresponding to 9 cycles in the mean span of 97.9 years is 10.88 years.

The 7 non-coherent cycles in the years 1798.3 to 1878.9 (SC5 through SC11) have a mean length of 11.51 years whereas the first 3 non-coherent Dalton Minimum sunspot cycles of 1798.3 to 1833.9 (SC5 through SC7) have a mean length of 11.87 years. The corresponding first 3 non-coherent sunspot cycles from 1976.1 to 2008.9 (SC21 through SC23) have a mean length of only 10.93 years.

To visualize the projection of the solar motion on the plane of the ecliptic during and shortly after the anomalous torque intervals, we used Carsten A. Arnholm's Solar Simulator 2 application for Figures 7-34. These snapshots also show ecliptic plane directions of selected planets and of the Sun's acceleration. Arnholm's +Y coordinate = $-Y_{ecliptic}$ and his +X = $-X_{ecliptic}$ (180$^o$ rotated). The red circle in Figures 7-34 is one solar diameter (1.392x10$^6$ km) from the barycenter.

All three paths in Figures 7-12 initially spiral in and then end with a small near-circular orbit, the center of which is quite offset from the barycenter. Two Sun, Jupiter, Saturn syzygies occur in each path. We categorize these three paths as phase restoration paths on strong heurist grounds, and find many surprising examples of planetary orbital resonance with the synodic period of Jupiter and Saturn in Figures 7-34. Mars synodic resonances are included in the charts, though Mars is not believed to have a significant role in sunspot formation.

The solar paths are all shown with respect to the fixed CRF/J2000.0 ecliptic plane. In the barycentric inertial frame, Jupiter and Saturn are currently progressing counterclockwise about 31$^o$ per 178.7 years, and there is a corresponding rotation of the barycentric solar path. The respective orbit inclinations to the ecliptic of Jupiter, Saturn, and Uranus are 1.31$^o$, 2.49$^o$, and 0.77$^o$. The inclination differences cause relatively small ± variations in the distance of the Sun from the ecliptic plane (see Figure 6), with a resulting slight distortion of the solar motion path projections on the ecliptic plane in Figures 7-34.

Other differences in the solar paths are caused by the Uranus torque cycle distortions and the Venus, Earth, and Mercury torque spikes of Figure 2 at the Jose Cycle intervals of interest. Even so, the similarities of the ecliptic plane projections of the solar paths in Figures 7-34 at various Jose Cycle time differences are striking.

The small but real solar path differences at Jose Cycle intervals may account for much, if not most, of the observed variation of ±0.9 year, 1 sigma from perfect Jose Cycle phase coherence of the observed sunspot cycle minima times in Table 1. The effects of Uranus' low frequency harmonic distortion and the high frequency inner planet torque spikes on the various solar paths also contribute to the observed variation of ± 2 years, 1 sigma for the observationally derived Jose Cycle times of sunspot maxima.

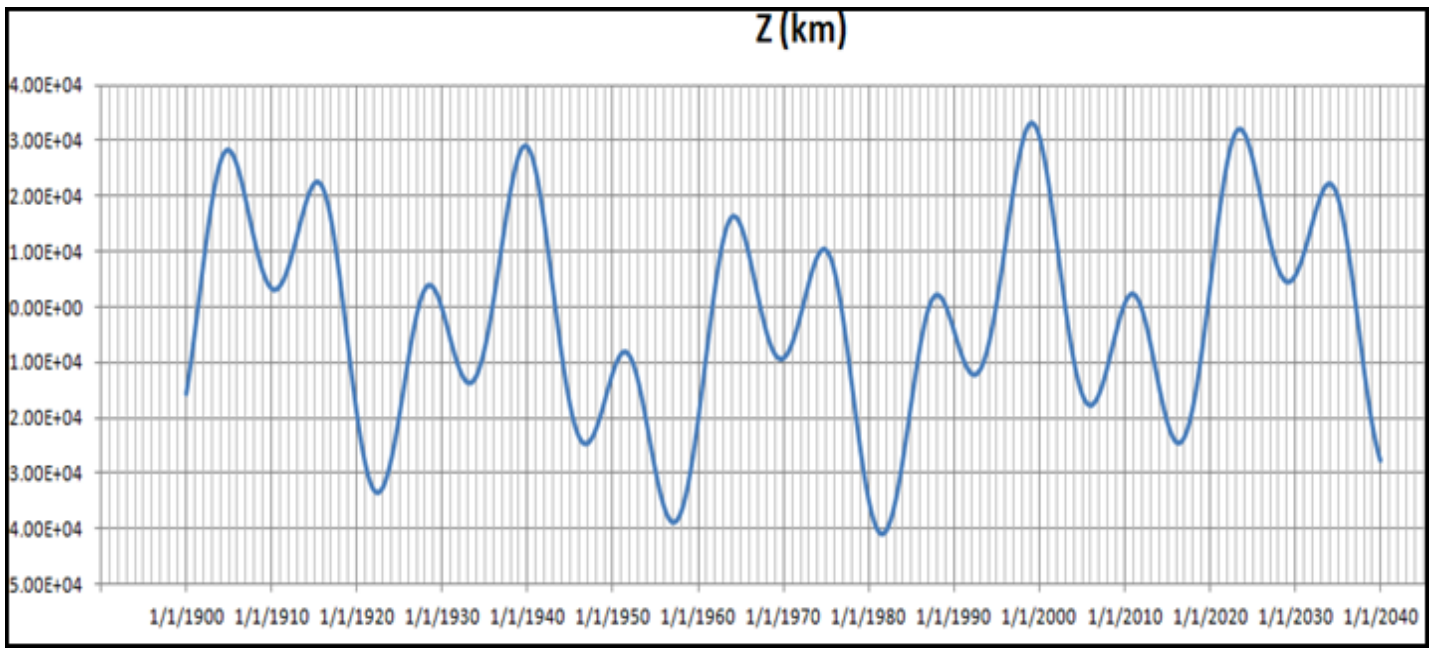

**Figure 6.** 1900-2040 ecliptic Z distances of the center of the Sun from the ICRF/J2000.0 ecliptic plane.

## 1.1 Phase Restoration Solar Paths (Figures 7-12)

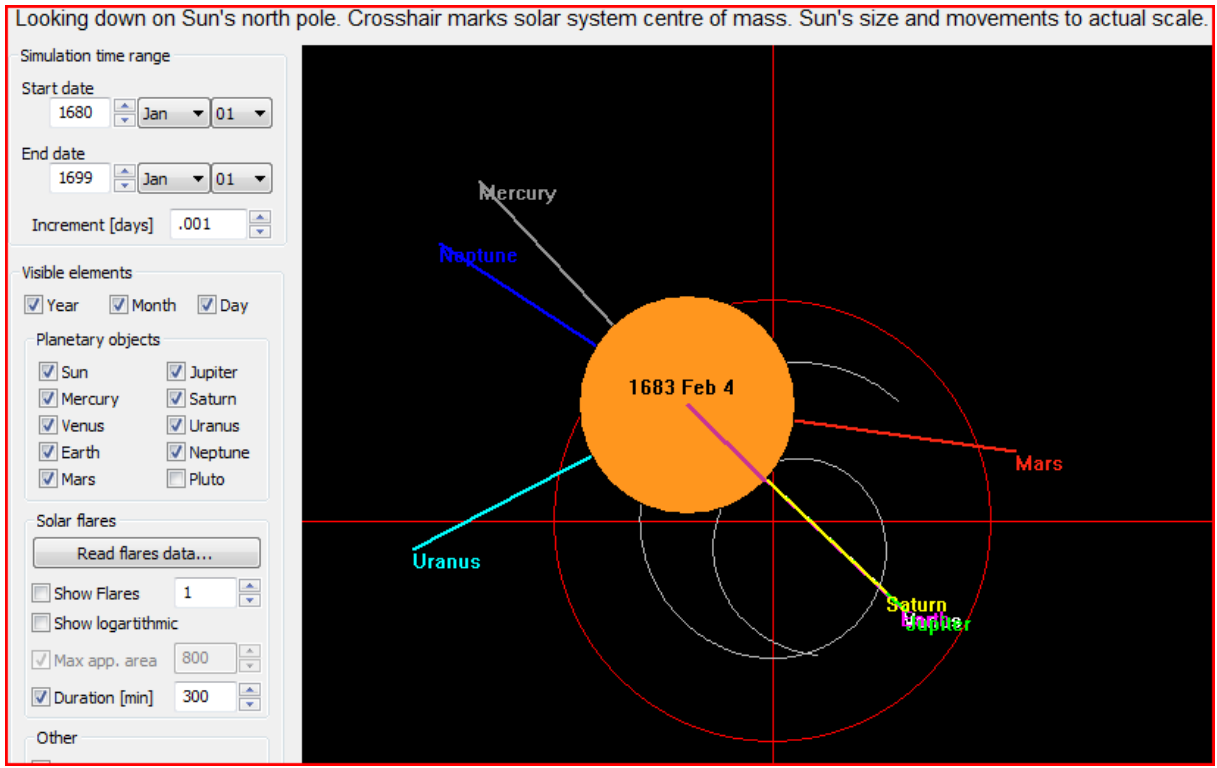

**Figure 7.** The phase restoration solar path from 1680 to 1699, showing a rare Sun, Venus, Earth, Jupiter, Saturn syzygy in 1683 (four-planet synodic resonance).

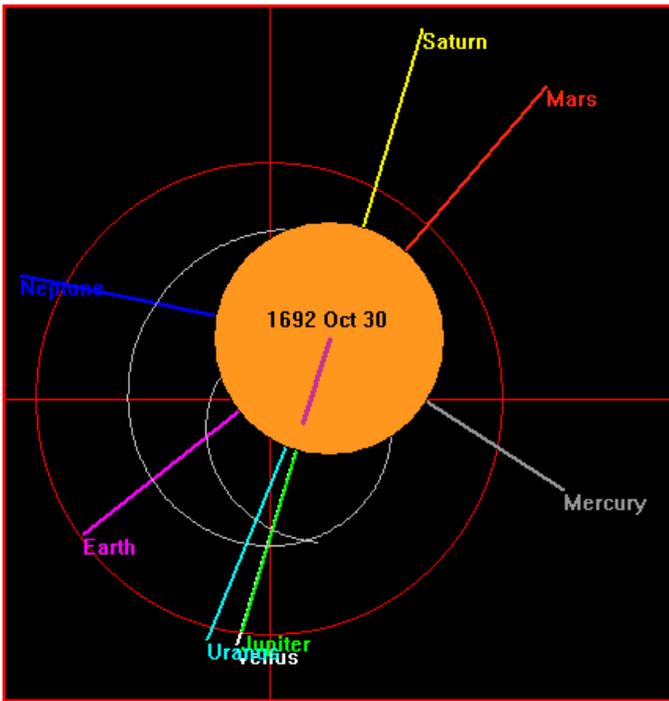

**Figure 8.** The phase restoration solar path from 1680 to 1699, showing the 1692 Sun, Venus, Jupiter, Saturn syzygy (three-planet resonance).

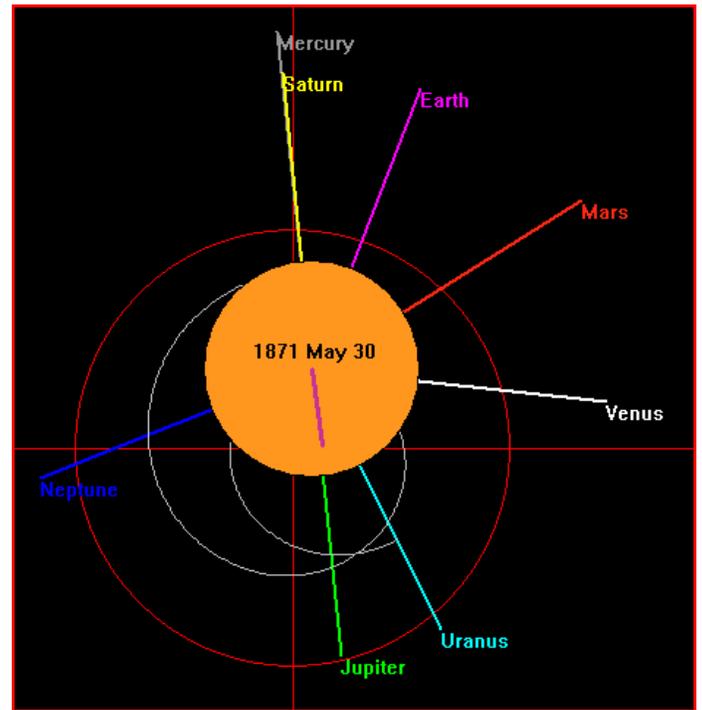

**Figure 10.** The phase restoration solar path from 1859 to 1878, showing the 1871 Sun, Mercury, Jupiter, Saturn (three-planet resonance) syzygy.

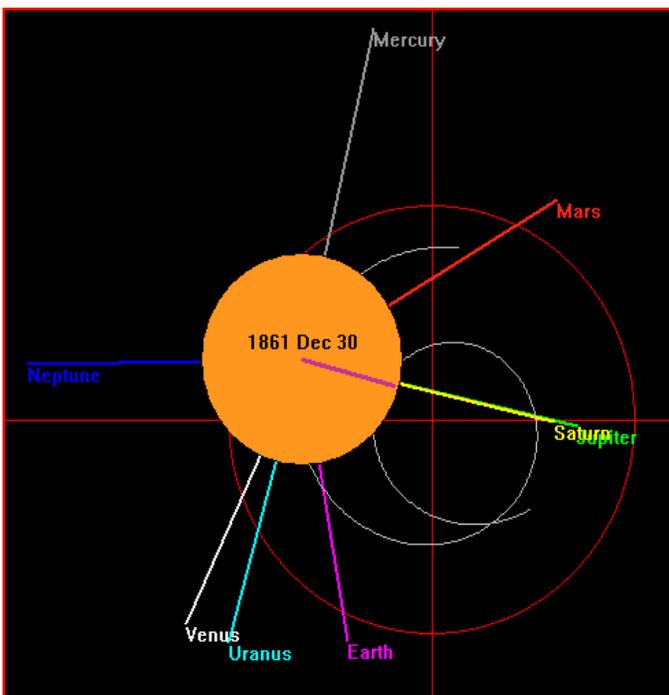

**Figure 9.** The phase restoration solar path from 1859 to 1878, showing the 1861 Sun, Jupiter, Saturn syzygy.

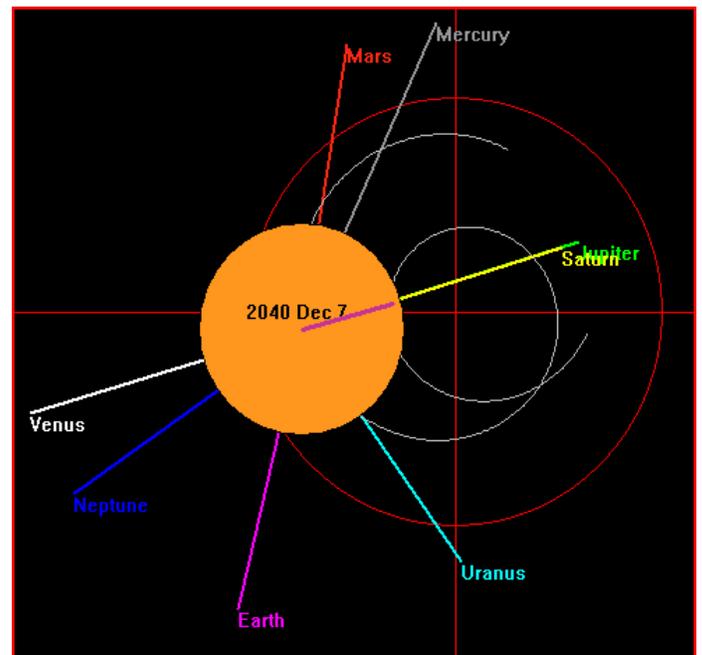

**Figure 11** The phase restoration solar path from 2036 to 2057, showing the 2040 Sun, Venus, Jupiter, Saturn syzygy (three-planet resonance).

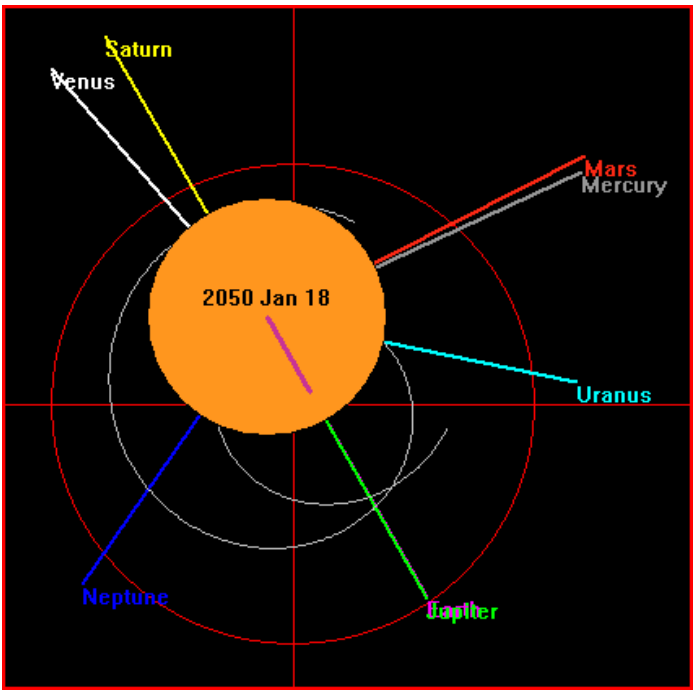

**Figure 12.** The phase restoration solar path from 2036 to 2057, showing the 2050 Sun, Earth, Jupiter Saturn syzygy (three-planet resonance).

Of the six Sun, Jupiter, Saturn syzygies in the above three sunspot cycle phase restoration paths, only two have associated sunspot cycle records from which we can make accurate phase comparisons. These are the 1862.0 and 1871.4 syzygies. The 1862.0 syzygy is 1.9 years past SC10 sunspot max and the 1871.4 syzygy is 0.5 years after SC11 sunspot max. The Jose cycle time differences for the three syzygy groups are shown in Table 2.

| J,S Syzygy | 1683.1 | 1692.83 | 1862.01 | 1871.42 | | |
|---|---|---|---|---|---|---|
| NextJose | 1862.01 | 1871.42 | 2040.93 | 2050.05 | Mean= | StdDev= |
| DerivedCycle | 178.91 | 178.59 | 178.92 | 178.63 | 178.76 | 0.183 |

**Table 2.** Derived Jose Cycle from Sun, Jupiter, Saturn syzygies in the 3 sunspot phase restoration solar paths.

**1.2 Sunspot Cycle Phase Disruptive Periods (Figures 13-18)**

The next six figures (Figures 13 - 18) show the solar paths during and immediately after the Figure 5 torque plateaus of 1609-1618, 1787-1796, and 1966-1976. In these solar paths, the sunspot cycles are separated by one Jose Cycle during the transition from phase coherent to non-coherent. We categorize them as sunspot cycle *phase disruptive* paths, aka "phase catastrophes" at the path ends. Whereas the sunspot cycle phase restoration solar paths of Figures 7-12 initially spiraled inward, ending with a small near-circular orbit with center offset from the barycenter, the sunspot cycle phase disruptive solar paths of Figures 13-18 begin with a small near-circular orbit with center offset from the barycenter, then all spiral outward to a maximum distance greater than one solar diameter from the solar system barycenter. The respective phase disruptive solar paths are 1609-1628 (Figures 13 and 14), 1787-1806 (Figures 15 and 16), and 1966-1985 (Figures 17 and 18). More examples of planetary synodic period resonance are seen in Figures 13-18. The signed torque signature for 1966-1985 can be seen in detail in that portion of Figure 2.

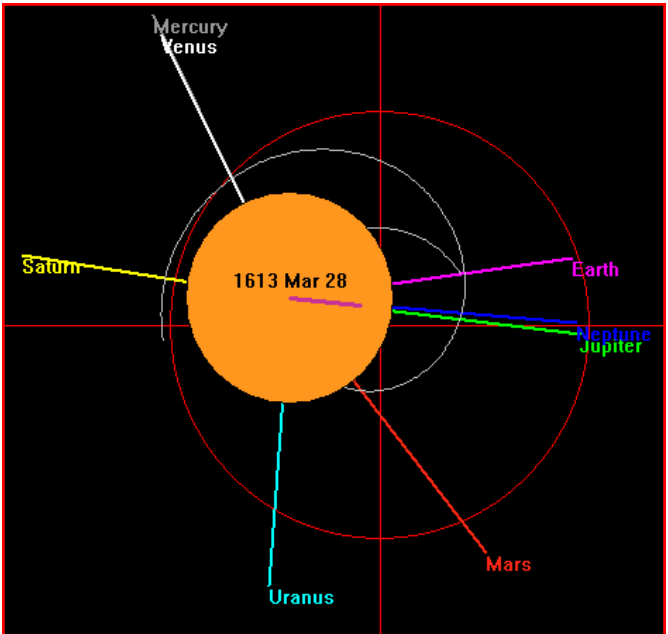

**Figure 13.** The phase disruptive solar path from 1609 to 1628, showing the 1613 Sun, Jupiter, Saturn and simultaneous Sun, Mercury, Venus syzygies (four-planet resonance).

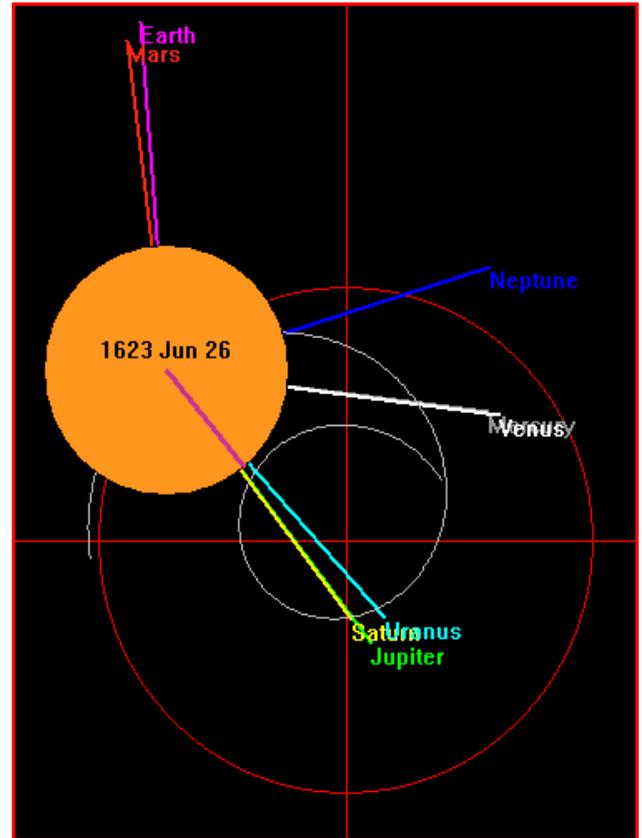

**Figure 14.** The phase disruptive solar path from 1609 to 1628, showing the 1623 Sun, Mercury, Venus and simultaneous Sun, Jupiter, Saturn syzygies (four-planet resonance).

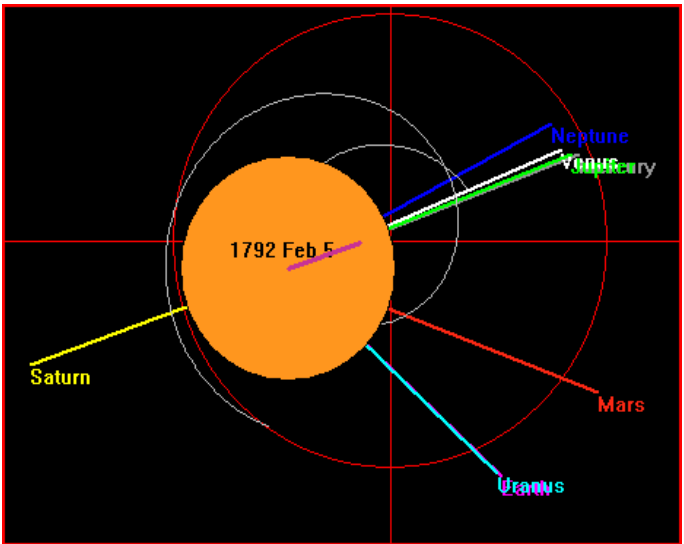

**Figure 15.** The phase disruptive solar path from 1787 to 1806, showing the 1792 Sun, Mercury, Jupiter, Saturn and simultaneous Sun, Earth, Uranus syzygies (five-planet resonance).

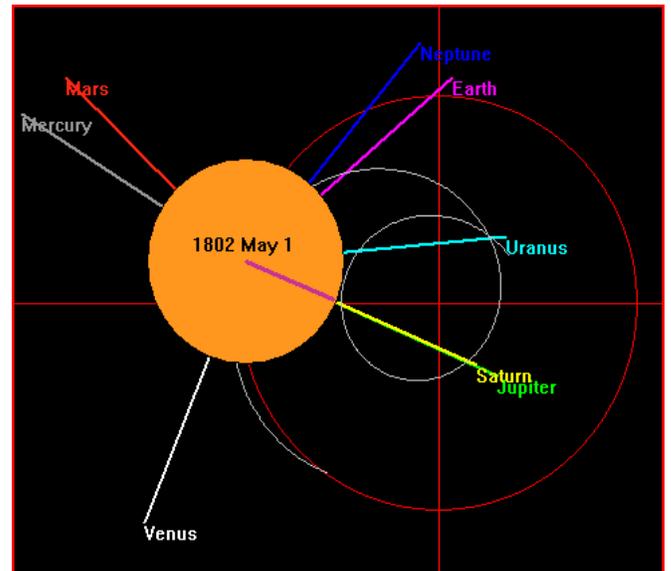

**Figure 16.** The phase disruptive solar path from 1787 to 1806, showing the 1802 Sun, Jupiter, Saturn syzygy.

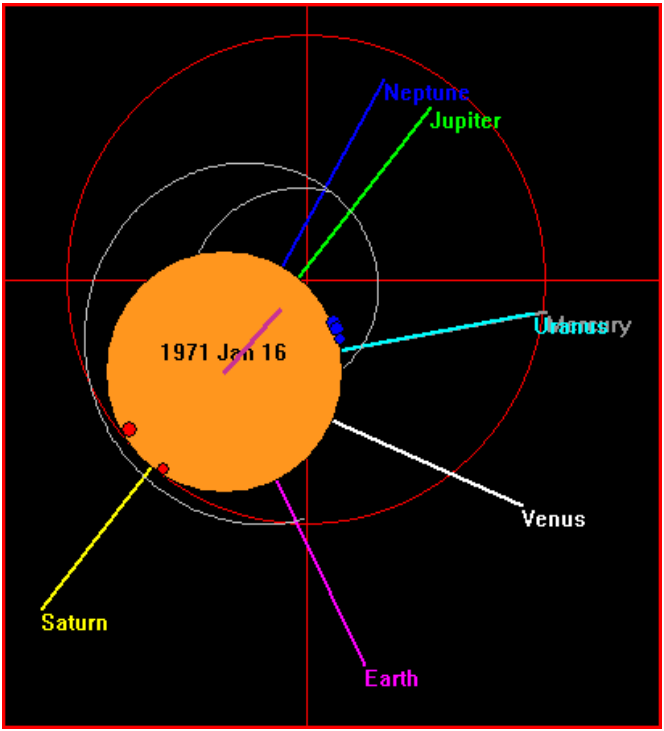

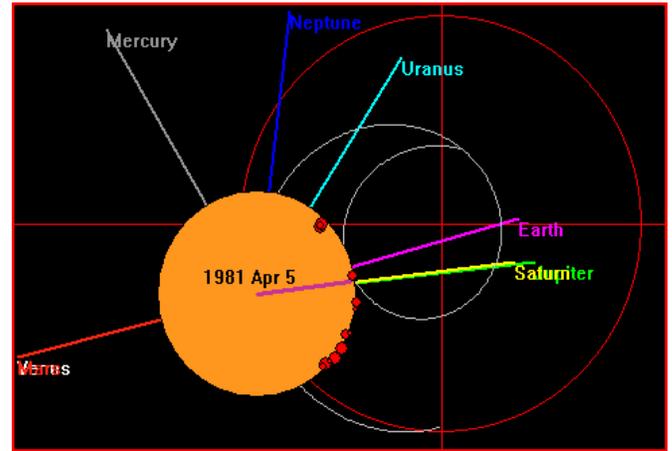

**Figure 18.** The phase disruptive solar path from 1966 to 1985, showing the 1981 Sun, Jupiter, Saturn syzygy, simultaneous Sun, Venus, Earth, Mars (five-planet synodic resonance) syzygy, and small solar flares on the Earth, Jupiter, and Saturn side of the Sun. As shown also in Figures 14 and 16, the Sun swings out beyond one solar diameter to offset the gravitational pull of the outer planets grouped on the other side of the barycenter.

Table 3 summarizes syzygy times and derived Jose Cycle lengths of Figures 13-18.

**Figure 17.** The phase disruptive solar path from 1966 to 1985, showing the 1971 Sun, Jupiter, Saturn and Sun, Mercury, Uranus syzygies (four-planet resonance) and solar flares in the direction of Saturn.

| J,S Syzygy | 1613.39 | 1623.49 | 1792.1 | 1802.33 | | |
| --- | --- | --- | --- | --- | --- | --- |
| NextJose | 1792.1 | 1802.33 | 1971.04 | 1981.26 | Mean = | StdDev= |
| DerivedCycle | 178.71 | 178.84 | 178.94 | 178.93 | 178.85 | 0.11 |

**Table 3.** Derived Jose Cycle from Sun, Jupiter, Saturn syzygies in the 3 sunspot cycle phase disruptive paths.

The 1792.095 Sun, Jupiter, Saturn syzygy of Figure 15 occurred near the middle of SC4 (1784.7-1798.3), 4.1 years past 1788 sunspot max. 178.94 years later, the 1971.044 Sun, Jupiter, Saturn syzygy occurred near the middle of SC20 (1964.8-1976.1), 1.8 years after 1969.2 sunspot max. The 1802.33 Sun, Jupiter, Saturn syzygy of Figure 16 occurred 4 years after the start of SC5(1798.3-1811), about 2.5 years before 1804.8 sunspot max. 178.93 years later, the 1981.26 Sun, Jupiter, Saturn and Sun, Venus, Earth, Mars syzygies occurred near the middle of SC21(1976.2-1986.6), about 1.4 years after 1979.9 sunspot max.

### 1.3 Detailed Comparison of Sunspot Non-coherent Periods (Figures 19-27)

At the time of writing this, we are still in a period of predominantly non-coherent sunspot cycle phase. We next examine the solar motion and sunspot cycles from 2006-2036, 1827-1859, and 1648-1680. Like the previous phase disruptive paths, these all start with a small near-circular orbit with center offset from the barycenter and swing out beyond one solar diameter from the barycenter, but these particular paths subsequently closely approach the barycenter and then spiral out to a tight loop, after which they will become the sunspot cycle phase restoration paths of Figures 7 through 12. The associated torque signature can be seen in detail for 2006-2036 in that portion of Figure 2.

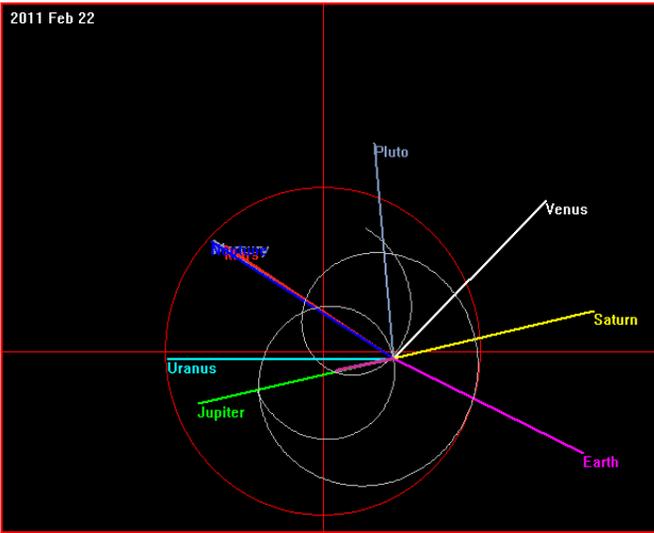

**Figure 19.** The solar path from 2006 to 2036, showing the Sun, Jupiter, Saturn syzygy and simultaneous Sun, Mercury, Mars, Neptune syzygy of 2011 (five-planet resonance).

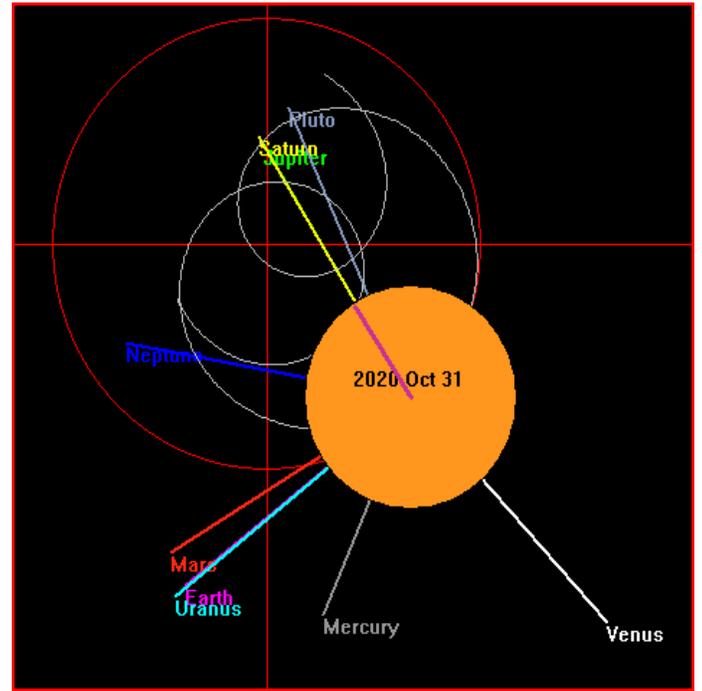

**Figure 20.** The solar path from 2006 to 2036, showing the Sun, Jupiter, Saturn syzygy of 2020.

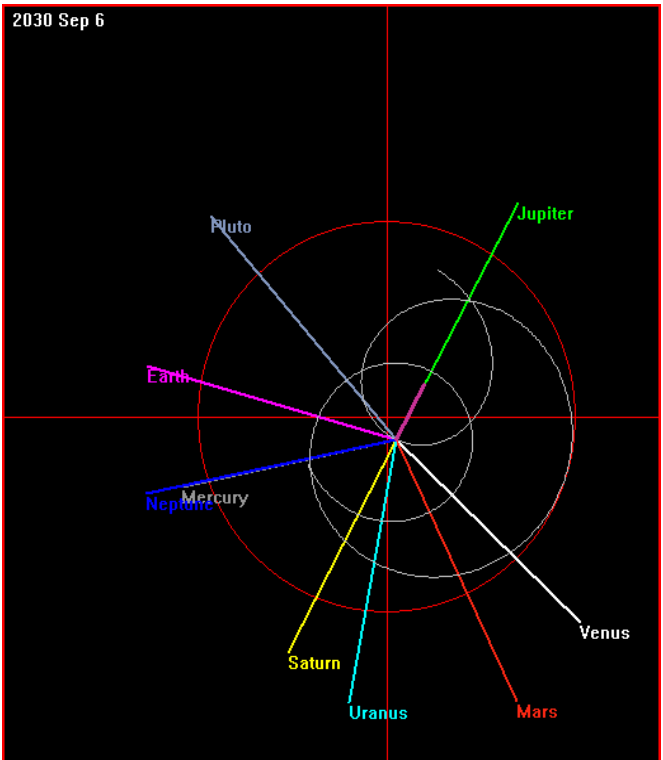

**Figure 21.** The solar path from 2006 to 2036, showing the Sun, Jupiter, Saturn and simultaneous Sun, Mercury, Neptune (four-planet resonance) syzygies of 2030 after a close solar flyby of the barycenter.

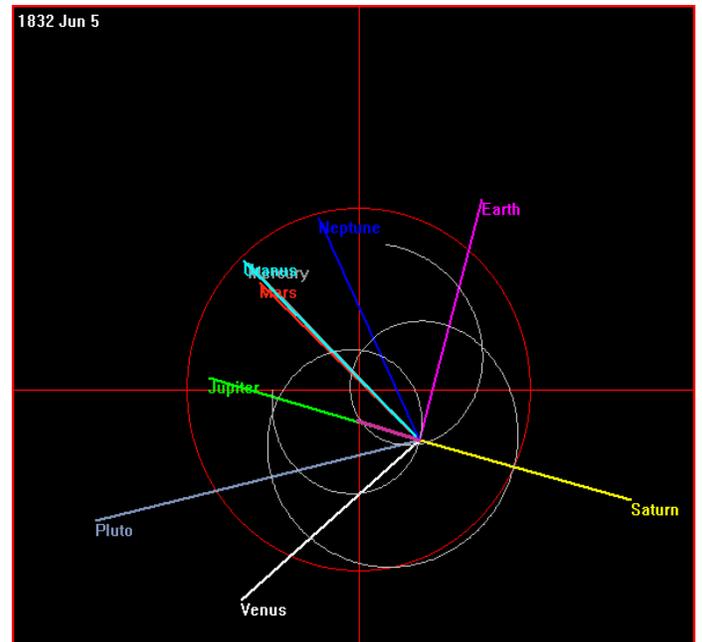

**Figure 22.** The solar path from 1827 to 1859, showing the Sun, Jupiter, Saturn and simultaneous Sun. Mercury, Uranus (four-planet resonance) syzygies of 1832.

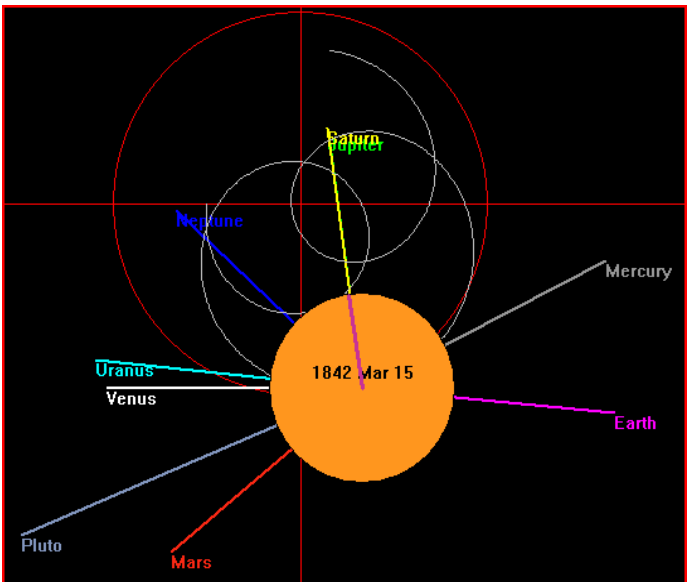

**Figure 23.** The solar path from 1827 to 1859, showing the Sun, Jupiter, Saturn and simultaneous Sun, Earth, Uranus syzygies of 1842 (third consecutive four-planet syzygy pair).

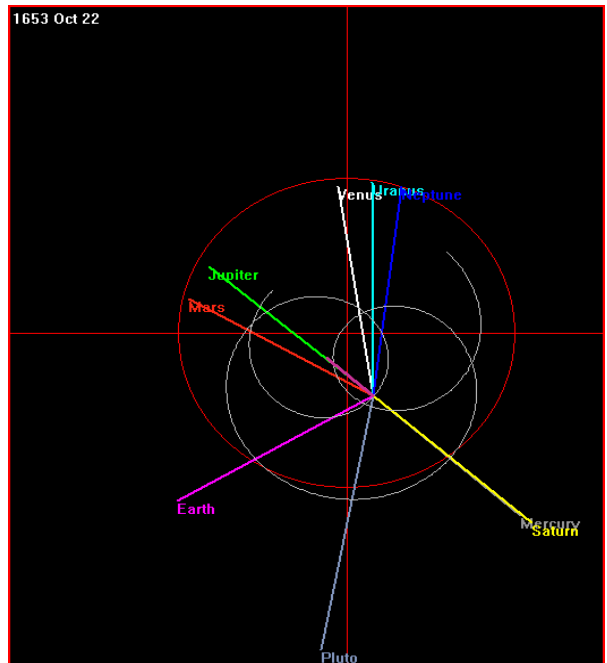

**Figure 25.** The solar path from 1648 to 1680, showing the Sun, Mercury, Jupiter, Saturn syzygy of 1653 (three-planet resonance).

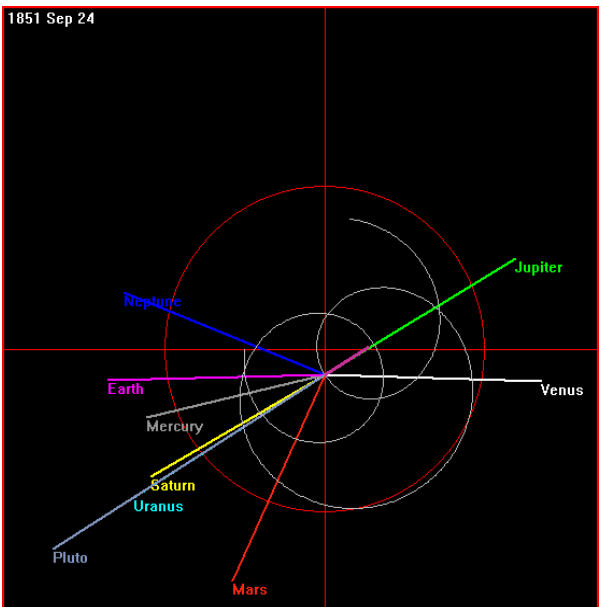

**Figure 24.** The solar path from 1827 to 1859, showing the Sun, Jupiter, Saturn and simultaneous Sun, Uranus, Pluto syzygies of 1851 (four-planet resonance) after a close solar flyby of the barycenter.

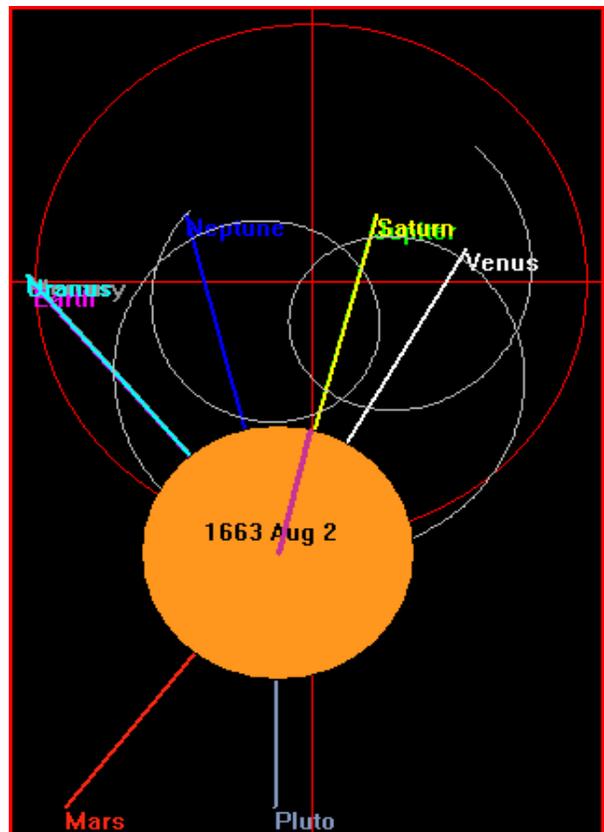

**Figure 26.** The solar path from 1648 to1680, showing the Sun, Jupiter, Saturn syzygy and simultaneous Sun, Earth, Uranus syzygy of 1663 four-planet resonance) with the Sun more than one solar diameter from the barycenter.

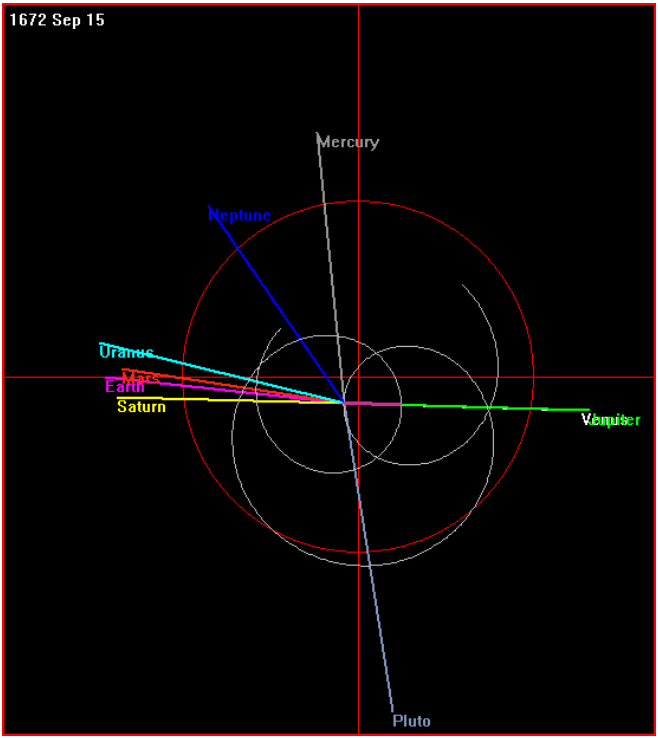

**Figure 27 (left).** The solar path from 1648 to 1680, showing the Sun, Venus, Jupiter, Saturn syzygy of 1672 (three planet resonance).

Distant dwarf planet Pluto was added in these figures for completeness only. Pluto's contributions to the torque cycles can be considered to be negligible.

| J, S Syzygy | 2011.145 | 2020.806 | 2030.682 | 1832.429 | 1842.203 | 1851.732 | | |
|---|---|---|---|---|---|---|---|---|
| Prev Jose | 1832.429 | 1842.203 | 1851.732 | 1653.808 | 1663.586 | 1672.708 | Mean= | Std Dev= |
| DerivedCycle | 178.716 | 178.603 | 178.95 | 178.621 | 178.617 | 179.024 | 178.755 | 0.185 |

**Table 4.** Derived Jose Cycle from Figures 19-27 Sun, Jupiter, Saturn syzygy times.

### 1.4 Detailed Comparison of Initial Sunspot Cycle Coherent Phase Solar Paths (Figures 28-34)

Next, please examine the solar paths encompassing the first two sunspot cycles in each of the sunspot phase coherent periods, 1699-1725, 1878-1904, and 2057-2083.

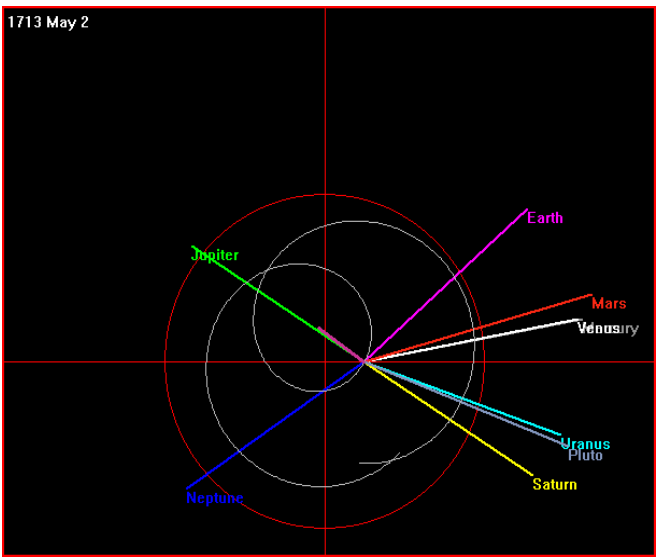

**Figure 28 (left).** The phase coherent solar path from 1699 to1725, showing the Sun, Jupiter, Saturn and simultaneous Sun, Mercury, Venus (four-planet resonance) syzygies of 1713.

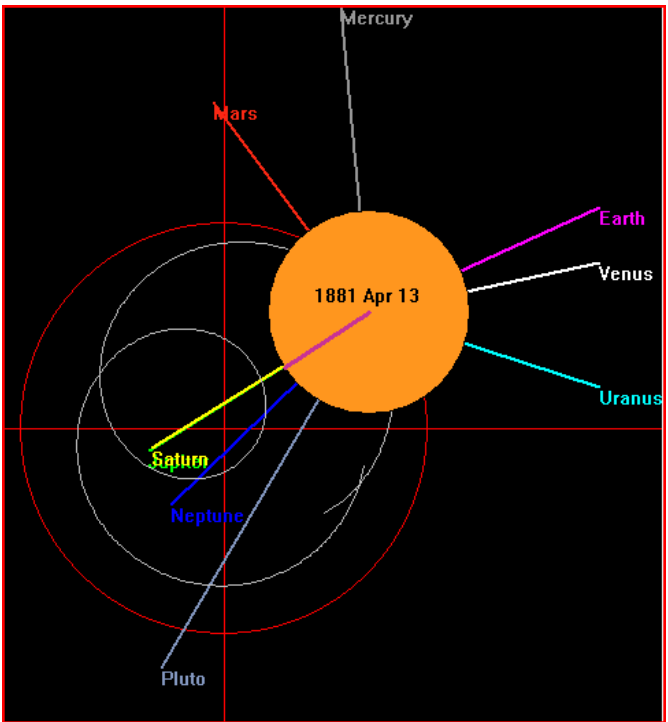

**Figure 29.** The phase coherent solar path from 1878 to 1904, showing the Sun, Jupiter, Saturn syzygy of 1881.

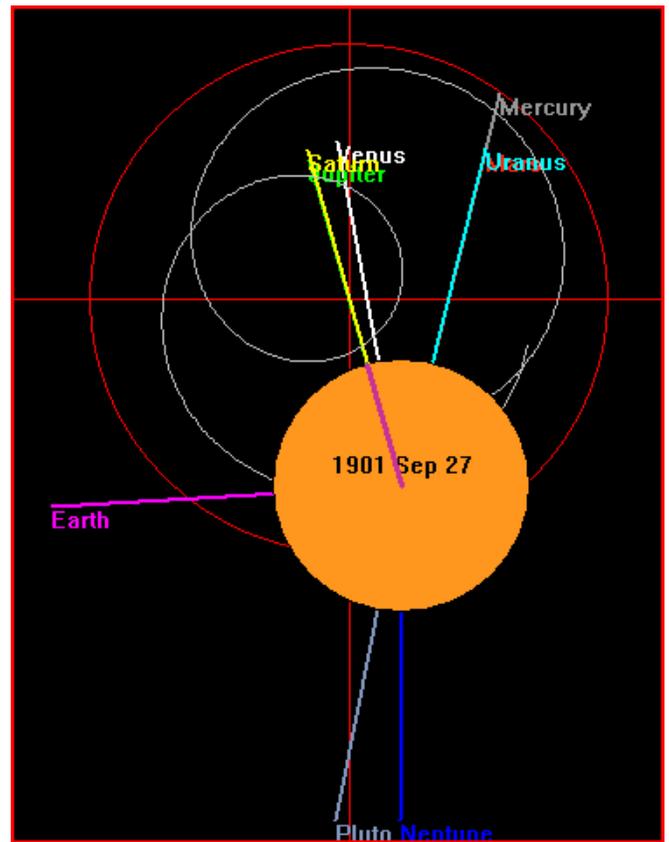

**Figure 31.** The phase coherent solar path from 1878 to 1904, showing the Sun, Jupiter, Saturn and simultaneous Sun, Mercury, Mars, Uranus (five-planet resonance) syzygies of 1901.

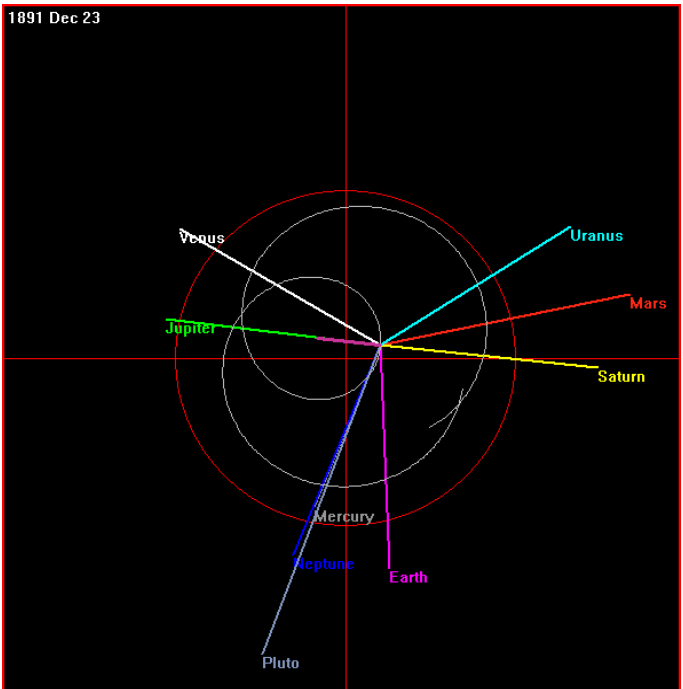

**Figure 30.** The phase coherent solar path from 1878 to 1904, showing the Sun, Jupiter, Saturn and simultaneous Sun, Mercury, Neptune (four-planet resonance) syzygies of 1891.

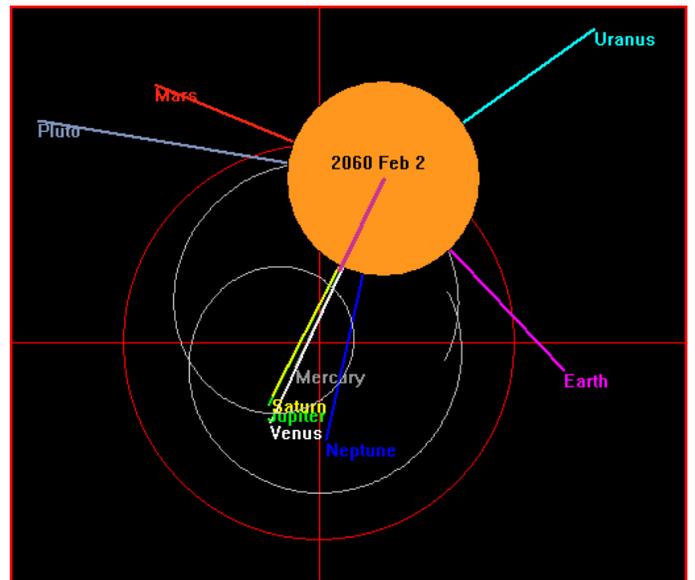

**Figure 32 (left).** The phase coherent solar path from 2057 to 2083, showing the Sun, Jupiter, Saturn and simultaneous Sun, Mercury, Venus (four-planet resonance) syzygies of 2060.

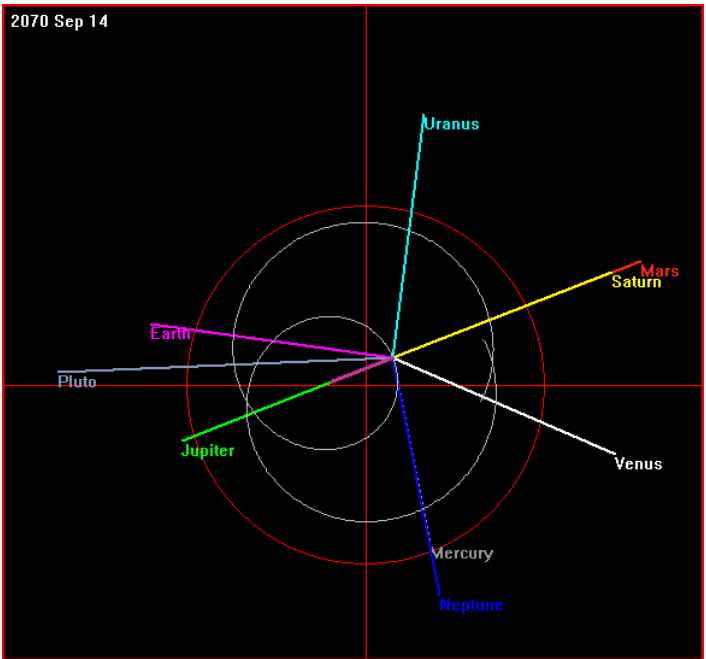

**Figure 33.** The phase coherent solar path from 2057 to 2083, showing the Sun, Mars, Jupiter, Saturn and simultaneous Sun, Mercury, Neptune (five-planet) syzygies of 2070.

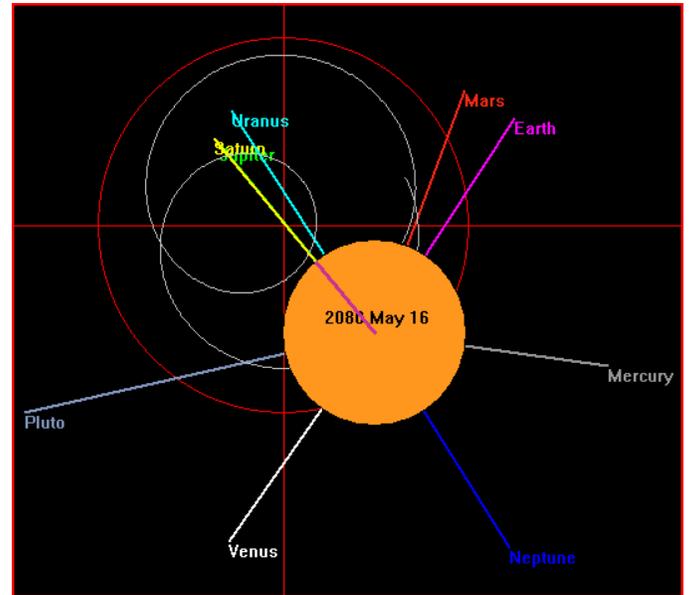

**Figure 34.** The phase coherent solar path from 2057 to 2083, showing the Sun, Jupiter, Saturn and simultaneous Sun, Venus, Earth (four-planet resonance) syzygies of 2080.

| J,S Syzygy | 2060.090 | 2070.704 | 2080.374 | | |
|---|---|---|---|---|---|
| Prev Jose | 1881.282 | 1891.978 | 1901.740 | Mean= | StdDev= |
| DerivedJose | 178.808 | 178.726 | 178.634 | 178.7227 | 0.087048 |

**Table 5.** Jose Cycle values derived from Sun, Jupiter, Saturn syzygy times in the solar paths of the projected phase coherent sunspot cycles SC28 and SC29.

**1.5 Detailed Planetary Synodic Resonances**

Table 6 shows the calculated Jose Cycle lengths corresponding to various planetary synodic pair resonances comprising a mean beat frequency of 5.58823 Jose cycles per millennium. The mean of the Table 6 derived cycle lengths is 178.94744 years ±0.1419 year, 1 sigma.

| PLANET | ORBIT PERIOD yrs | SYNODIC | WITH | 1/(1/S1-1/S2) | X | Jose Cycle |
|---|---|---|---|---|---|---|
| NEPTUNE | 164.7858 | JUPITER | NEPTUNE | 12.78211217 | 14 | 178.94957 |
| SATURN | 29.457 | JUPITER | SATURN | 19.85899028 | 9 | 178.73091 |
| MARS | 1.8808 | MARS | JUPITER | 2.235207149 | 80 | 178.81657 |
| VENUS | 0.6152 | VENUS | EARTH | 1.598752599 | 112 | 179.06029 |
| EARTH | 1 | EARTH | JUPITER | 1.092064077 | 164 | 179.09851 |
| JUPITER | 11.862 | VENUS | JUPITER | 0.648851442 | 276 | 179.0830 |
| MERCURY | 0.24084 | MERCURY | VENUS | 0.395781515 | 452 | 178.89324 |

**Table 6.** Synodic planetary resonances at the Jose Cycle frequency

Not included in Table 6, Uranus has a sidereal orbital period of 84.011 years. The Jupiter, Uranus synodic resonance closest to 179 years has a 4-sigma outlier value of 179.559 years. Figure 8 shows an example of the small offset of Uranus from synodic resonance with Jupiter, Saturn, and Venus on October 30, 1692. Figure 22 shows an instance of Mercury resonance with Uranus on June 5, 1832.

A resonant Sun, Earth, Uranus syzygy on March 15, 1842 is also shown in Figure 23, and Figure 24 shows a synodic Pluto resonance with Uranus on September 24, 1851. These were all in the sunspot cycle predominantly non-coherent period of 1827-1859, but a Sun, Mercury, Mars, Uranus syzygy also occurred on Sept 27, 1901 in a period of sunspot cycle phase coherency (see Figure 31).

**2.0 Predicted Future Solar Cycle Phase Coherence, SC28 to SC36**

Looking ahead to 2057 and beyond, we confidently predict that sunspot cycles SC28 through SC35 (2057.5 to 2143.5) will be phase coherent at times of minima and amplitude correlated at maxima with SC12 through SC19 (1878.8 to 1964.5). The predicted start times (±0.9 year, 1 sigma) of SC28 through SC36 are tabulated in Table 7. We expect sunspot phase coherence analysis capability to be significantly refined before 2057, with consequent smaller uncertainties associated with long term predicted sunspot cycle start times, even though these uncertainties are already much less than for any other presently known long term prediction methods.

| Sunspot Cycle | SC28 | SC29 | SC30 | SC31 | SC32 | SC33 | SC34 | SC35 | SC36 |
|---|---|---|---|---|---|---|---|---|---|
| Coherent with: | SC12 | SC13 | SC14 | SC15 | SC16 | SC17 | SC18 | SC19 | SC20 |
| Start Year | 1878.8 | 1890.2 | 1902.0 | 1913.6 | 1923.6 | 1933.7 | 1944.3 | 1954.3 | 1964.8 |
| Start+178.7years | 2057.5 | 2068.9 | 2080.7 | 2092.3 | 2102.3 | 2112.4 | 2123.0 | 2133.0 | 2143.5 |

Table 7. Predicted Sunspot Cycle Start Times ±0.9 year, 1 sigma, using the 178.7 year coherence cycle from Table 5 and standard deviation based on the Table 1 observed phase jitter.

**3.0 Conclusions and Outlook**

We have verified Paul D. Jose's discovery that the torque cycles induced on the Sun by planetary perturbations repeat at ~179 year intervals with a high degree of phase and amplitude coherence. We have also identified and examined two 98 year Jose subcycles of consistent sunspot cycle phase coherence at times of minima and amplitude correlation of maxima. These are SC12 through SC20, 1878.8-1976.1 with SC-[5] through SC4, 1699-1798.8), in which the solar paths remain within one solar diameter distance from the barycenter.

Also identified were three 81 year subcycles of predominantly non-coherent sunspot cycle phasing (SC21 through SC27, 1976-2057, SC5 through SC11, 1798.3-1878.8, and 1618-1699), in which the solar paths had strayed well beyond one solar diameter distance from the barycenter in the phase disruptive segments.

These subcycles alternate and repeat in the 98+81 = 179 year Jose Cycle as the Sun moves in a complex but extremely predictable path that continuously repeats at the Jose frequency. The observed Jose Cycle phase coherence at times of minima of the 9 Schwabe sunspot cycles with the mean 98 year subcycle, combined with the stable phase restoration solar paths of 2036-2057 (Figure 11) and 2057-2083 (Figure 32), enable us to heuristically predict the start times of future sunspot cycles SC28 to SC36 (2057.5-2143.5) with an associated uncertainty level of ± 0.9 year, 1 sigma, corresponding to the Table 1 observed phase jitter.

At present we are not aware of any other method capable of predicting the start times of future sunspot cycles as well as this for even as little as two cycles in advance. It is also anticipated that sunspot activity will increase over the period 2057-2136 in a pattern similar to the increasing activity in the periods 1699-1778 and 1878-1957 shown in Figure 4.

The actual magnitudes of SC28 to SC36 will likely be similar to those of SC12 to SC20, starting low and increasing to similarly high peak magnitudes. In the course of our investigation we also made a compilation of previously unrecognized planetary synodic resonances driven by Jupiter at the Jose Cycle beat frequency. The much higher synodic resonance frequencies of the inner planets Mercury, Venus, and Earth are observed to systematically augment the Jose beat frequency. This is additional evidence that Venus, Earth, and Mercury directly reinforce Jupiter and Saturn in the production of the observed Schwabe sunspot cycles.

We are confident that an astrophysical theory of the solar system sunspot dynamo (SSSD) that is mathematically rigorous will eventually emerge, but for that to happen a serious examination of the points of intersection of celestial mechanics and astrophysics will probably be necessary. These have traditionally been separate specialized disciplines.

A controversial hypothesis that seems consistent with our empirical findings is presented in the 2008 paper, *Does a Spin–Orbit Coupling Between the Sun and the Jovian Planets Govern the Solar Cycle?* by I. R. G. Wilson, B. D. Carter, and I. A. Waite. Interestingly, the authors seemed aware of the importance of the Jose Cycle but, like Paul Jose himself, were unable to specifically identify the three crucial 81 year subcycles of predominantly non-coherent sunspot cycles and the two 98 year subcycles of coherent sunspot cycle activity. Though they admitted to not having a rigorous physical SSSD theory, they did show why the empirical evidence suggests a possible spin-orbit coupling process.

F. Stefani, A. Giesecke, N. Weber, and T. Weier have concluded that very little, if any, energy transfer from planets to the Sun is required to produce toroidal-to-poloidal field helicity oscillations recently found in simulations of a current-driven, kink-type Tayler instability.

Another conceivable mechanism of sunspot formation is magnetic modulation by the strong magnetic fields of Jupiter and Saturn through periodic flux transfer to the Sun's tachocline. At present, there does not seem to be observational evidence of periodic flux transfer, but that may simply be because nobody has proposed a viable observation program for its detection.

Any observed effects of Jupiter and Saturn magnetic modulation would have to be traceable from Saturn and Jupiter back to the Sun's tachocline. A single flux transfer event on the dayside of Saturn's magnetosphere has been observed by the Cassini spacecraft, and FTEs that intermittently connect Earth's magnetic field to the Sun's through field-aligned currents have been observed by THEMIS and ESA Cluster satellites. Planetary magnetic field sunspot modulation is a promising area for further investigation.

# References

Arnholm, C. A., Solar Simulator 2, [www.arnholm.org/astro/sun/sc24/sim2/](www.arnholm.org/astro/sun/sc24/sim2/)

Eddy, J. A. 1976, The Maunder Minimum, Science, Vol. 192, p. 1189

Jasinski, J. M., Slavin, J. A., Arridge, C. S., Poh, G., Jia, X., Sergis, N., Coates, A. J., Jones, G. H., Waite, J. H., Flux transfer event observation at Saturn's dayside magnetopause by the Cassini spacecraft, 6 July, 2016, Geophysical Research Letters, [onlinelibrary.wiley.com/doi/10.1002/2016GL069260/full](onlinelibrary.wiley.com/doi/10.1002/2016GL069260/full)

Jose, P. D., Sun's Motion and Sunspots, 1965, AJ, 70, 193

Klepper, D., Angular Momentum, [http://web.mit.edu/8.01t/www/materials/modules/chapter19.pdf](http://web.mit.edu/8.01t/www/materials/modules/chapter19.pdf)

Stefani, F., Giesecke, A., Weber, N., Weier, T., Synchronized Helicity Oscillations: A Link Between Planetary Tides and the Solar Cycle?  arXiv:1511.09335, Solar Physics, V2, 5 Aug 2016

Sunspot Index and Longterm Solar Observations, [www.sidc.be/silso/versionarchive](www.sidc.be/silso/versionarchive), Version 1.0

Svalgaard, L., Cagnotti, M., Cortesi, S., The Effect of Sunspot Weighting, [https://arxiv.org/ftp/arxiv/papers/1507/1507.01119.pdf](https://arxiv.org/ftp/arxiv/papers/1507/1507.01119.pdf)

Thomas, B. T., Smith, E. J., The structure and dynamics of the heliospheric current sheet, J. Geophysical Research: Space Physics (1978–2012), [Volume 86, Issue A13,](#) pages 11105–11110, 1 December 1981

WDC-SILSO sunspot data, Version 2 of data set, [www.solen.info/solar/cycles1_to_present](www.solen.info/solar/cycles1_to_present)

Wilson, I. R. G., Carter, B. D., Waite I. A., Does a Spin–Orbit Coupling Between the Sun and the Jovian Planets Govern the Solar Cycle?  Astronomical Society of Australia, Vol. 25, No. 2, 2008, pp. 85-93